\documentclass[conference]{IEEEtran}
\usepackage{cite}
\usepackage{amsmath,amssymb,amsfonts}
\usepackage{algorithmic}
\usepackage{graphicx}
\usepackage{textcomp}
\usepackage{xcolor}
\usepackage{url}
\usepackage{hyperref} 
\def\BibTeX{{\rm B\kern-.05em{\sc i\kern-.025em b}\kern-.08em
    T\kern-.1667em\lower.7ex\hbox{E}\kern-.125emX}}
\begin{document}

\title{AI Agent Communications in AI-Native 6G Network: Status, Challenges and Opportunities
}

\author{\IEEEauthorblockN{Qiang Duan}
\IEEEauthorblockA{\textit{Information Sciences \& Technology Department} \\
\textit{Pennsylvania State University}\\
Abington, USA \\
qduan@psu.edu}
}

\maketitle

\begin{abstract}
The rapid development of agentic AI and multi-agent systems is establishing AI agent communication as a fundamental requirement for the future Internet. While a diverse array of agent communication protocols has recently emerged, these solutions currently suffer from interoperability crises and infrastructure gaps. The newly proposed Service-Oriented Virtualization-Based Architecture (SOVA) offers an architectural framework to address these challenges for agent communication, which expects seamless support from the network infrastructure. The emerging AI-native 6G network is promising as a robust foundation for the SOVA framework, thereby greatly facilitating AI agent communication; however, its effectiveness in supporting the SOVA framework has yet to be fully assessed. To bridge the distinct research trajectories of AI-native 6G networks and AI agent communications, this paper investigates the capabilities of current and proposed 6G network architectures and protocol specifications for supporting the SOVA framework for AI agent communications. By critically examining 6G's key architectural paradigms and their potential to fulfill SOVA's requirements, this paper identifies gaps between 6G standards and the demands of AI agent communication. Based on this gap analysis, this paper outlines research and development directions to ensure that the future 6G network can natively empower AI agent communications in the era of agentic AI. 
\end{abstract}


\section{Introduction}

The recent rapid development in agentic AI in general, and multi-agent systems (MAS) in particular, is making communication among autonomous LLM-based AI agents a crucial demand for the future Internet \cite{Rise-Agent-survey-2025}, \cite{Agentic-AI-2026}, \cite{Self-talk-2025}. The latest technical progress, such as the NVIDIA RTX Spark system for AI PCs, is enabling agentic edge intelligence, with numerous AI agents deployable on local devices. The demand for communications among these agentic AI devices will trigger a new wave of the Internet of AI agents, in which agent communication protocols play a critical role \cite{duan2025agent}.

A wide spectrum of research efforts have been undertaken to develop agent communication protocols that support the emergence of the Internet of AI Agents. A proliferation of such protocols has been proposed, each designed to optimize distinct operational environments \cite{Agent-Interoper-Survey-2025}, \cite{Agent-Comm-Survey-2025}. However, current agent communication protocols suffer from an interoperability crisis and an infrastructure gap, operating largely independently of the networks that host them. To harmonize these diverse communication protocols and support massive, internet-scale MAS deployments, researchers have exploited the principles of virtualization and service-orientation and proposed a Service-Oriented Virtualization-Based Architecture (SOVA) framework \cite{duan2026agent}, which abstracts heterogeneous infrastructure resources and AI agent capabilities into scalable, composable service components, thereby facilitating seamless cross-domain interoperability.  

For the Internet of AI agents to function efficiently, the underlying network infrastructure must evolve from a passive conduit of data into an intelligent, proactive participant in information processing. The forthcoming 6G network is being conceptualized and standardized precisely to fulfill this role. AI-nativeness is expected to be a key attribute of the future 6G network. In the AI-native 6G network, agentic AI becomes an indispensable component of the network system, and the network provides a platform to support agentic AI through an AI-as-a-service (AIaaS) paradigm \cite{AI-Comm-6G-25}. The principles of virtualization and service-oriented architecture are fully embraced by the 6G network in both architectural design and protocol specifications, for example, through network slicing and the service-based architecture (SBA). Therefore, the 6G network offers a robust foundation for implementing the SOVA framework for AI agent communications toward enabling the Internet of AI Agents.  
However, realizing this synergy between the AI-native 6G network and AI agent communications requires overcoming profound challenges. 6G standardization efforts, while visionary in their inclusion of AI entities, have yet to fully articulate the protocol-level mechanics required to support dynamic, multi-agent interactions across heterogeneous domains. 

To bridge the distinct research trajectories of AI-native 6G networks and AI agent communications, this paper investigates the capabilities of current and proposed 6G network architectures and protocol specifications in supporting the SOVA framework for AI agent communications. By critically examining 6G's Service-Based Architecture (SBA), its mechanisms for network virtualization and slicing, and its foundational AI-native design principles, this paper identifies gaps between 6G standards and the requirements of AI agent communication. Based on this gap analysis, this paper delineates a research and development roadmap designed to ensure that future 6G networks can natively empower the Internet of AI agents, transforming AI agent communication into a seamless, intrinsic feature of global telecommunications infrastructure.

\section{Agent Communications for Agentic AI}

\subsection{Current Agent Communication Protocols and Limitations}

The current landscape of AI agent communication protocols features diverse solutions tailored to specific operational environments and requirements. Representative protocols for inter-agent communications include A2A, ACP, ANP, LMOS, and AConP.

The Agent2Agent (A2A) protocol uses a client-server model based on JSON-RPC 2.0, with a core design principle of ``opaque execution'' that enables secure enterprise collaboration without exposing internal agent memory states or prompts \cite{A2A-Review-2025}, \cite{Secure-A2A-2025}. It supports both synchronous and asynchronous modes, utilizing Server-Sent Events and Webhooks to manage complex tasks. To combat fragmentation, A2A recently merged with the Agent Communication Protocol (ACP) \cite{ACP-2026}—a REST-based framework initially designed for long-running, asynchronous workflows—establishing a unified standard under the Linux Foundation.  

For decentralized environments, the Agent Network Protocol (ANP) acts as the ``HTTP of the agentic web.'' \cite{ANP-WP-2025} Using a peer-to-peer architecture, ANP employs W3C Decentralized Identifiers (DIDs) \cite{DID-survey-2025} and JSON-LD \cite{JSON-LD-spec} to enable trustless, cross-organizational agent discovery and communication over open networks.  

The Language Model Operating System (LMOS) provides a cloud-native framework for multi-agent ecosystems \cite{LMOS}. Built on the W3C Web of Things (WoT) standard, LMOS uses JSON-LD Thing Descriptions to enable protocol-agnostic interactions, allowing agents to seamlessly switch between HTTP, MQTT \cite{MQTT-spec}, and CoAP \cite{CoAP-spec} depending on infrastructure needs.  

The Agent Connect Protocol (AConP) within the AGNTCY project  \cite{AGNTCY-origins} exploits gRPC-based binary Protobuf serialization to reduce payload sizes and achieve short response times, making it ideal for real-time vehicular networks and resource-constrained edge deployments \cite{Performance-gRPC-2024}.

The proliferation of agent protocol designs has led to the coexistence of competing protocols, thereby introducing interoperability issues on the communication level. For example, agents running A2A in an enterprise environment cannot directly interact with agents running ANP on a public network. 
The current trend in agent protocol development suggests that no single protocol is likely to dominate agent communication across all agentic AI application scenarios. Analysis also indicates that no single technology can be optimized to address all the conflicting requirements of agent communication across diverse scenarios \cite{duan2026agent}. Therefore, there is an interoperability crisis among diverse agent protocols tailored to various networking scenarios.  

Although agent protocols attempt to enhance infrastructure awareness and provide various cloud-centric or edge-native designs, there still lacks an effective mechanism that enables smooth cooperation between resource-aware management at the agent communication layer and the control mechanisms in the infrastructures, e.g., task scheduling in edge computing systems, bandwidth allocation in networks, and virtual machine scaling in cloud data centers. Therefore, the infrastructure gap of AI agent communication remains unfilled. 

\subsection{Service-Oriented Virtualization for Agent Communications}

\begin{figure} 
\centering
\includegraphics[width=3.45in]{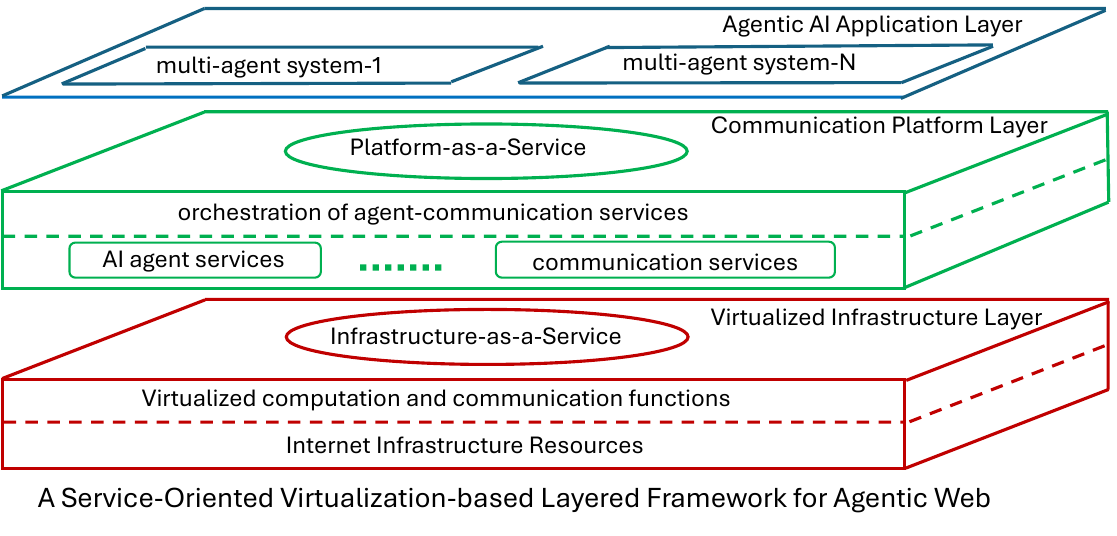}
\caption{{The service-oriented virtualization-based architectural (SOVA) framework for AI agent communication \cite{duan2026agent}.}}
\label{SOVA}
\end{figure}

To resolve the interoperability crisis and bridge the infrastructure gap in current agent protocols, we proposed the Service-Oriented Virtualization-Based Architecture (SOVA) in our previous work \cite{duan2026agent}, as depicted in Figure \ref{SOVA}. Through a pervasive application of the Everything-as-a-Service (XaaS) paradigm, the SOVA framework constructs a layered architecture comprising the Virtualized Infrastructure layer at the bottom, the MAS Communication Platform layer in the middle, and the Agentic AI Application layer at the top. 

The SOVA framework applies the proven strategies of virtualization and service orientation to the domain of multi-agent communication. Virtualization allows multiple isolated multi-agent systems to share a common physical edge-network-cloud substrate without causing resource contention or data interference. The service-oriented architecture of SOVA enables effective cross-layer interaction through the IaaS/PaaS interfaces and flexible cross-domain cooperation through service orchestration. Therefore, SOVA offers a unified architectural framework that not only accommodates the coexistence of hybrid agent protocols with diverse design objectives but also adaptively leverages the appropriate protocol bundle to support the numerous MASs deployed for various agentic AI applications.

\section{AI-Native 6G Network for AI Agent Communications}


The evolution from 5G to 6G is characterized not merely by an anticipated increase in radio spectrum bandwidth but by a fundamental architectural paradigm shift toward true AI-nativeness. 
AI is no longer treated as an external, over-the-top application on the user plane; it is being natively integrated into the network's foundational control-plane fabric \cite{AI-native-6G-review-2025}.

This section investigates how the AI-native 6G network may support the Internet of AI agents by identifying the most relevant attributes of the emerging 6G network and analyzing how each attribute facilitates the realization of the SOVA framework for AI agent communication. 

\subsection{Advanced Network Slicing and Compute-Network Convergence} 

Network slicing in 6G advances from a rudimentary network overlay concept into an intelligent, multidimensional resource allocation mechanism. In the 5G era, network slicing primarily partitioned bandwidth and latency parameters to support broad, generalized service categories such as Ultra-Reliable Low-Latency Communication (uRLLC) and enhanced Mobile Broadband (eMBB) \cite{NetSlice-5G-2019}. In the 6G network, slicing capabilities become highly recursive, deeply programmable, and infinitely granular \cite{NetSlice-6G-2022}, \cite{AI-Comm-6G-25}. Furthermore, the 6G network departs from the conventional ``cloud computing plus communication network '' model and embeds massive computational resources directly into the network fabric. This profound architectural shift is formally recognized as Computing-Network Convergence (CNC) \cite{duan2020convergence}, \cite{CNC-6G-26}. 

Because AI agents demand significant, often bursty, computational resources for model inference alongside data transmission, 6G network slices allocate combined communication bandwidth, edge caching, and distributed computational power simultaneously \cite{Joint-NS-24}. This capability directly fulfills SOVA’s virtualization requirement. Through Computing-Networking Convergence, 6G orchestration mechanisms, such as advanced iterations of the Network Slice Management Function (NSMF), can dynamically allocate and isolate not only radio frequencies and transport bandwidth but also the precise edge computing resources required for heavy AI inference workloads \cite{NS-VNF-26}; therefore, greatly facilitating communications for multi-agent collaborations.

\subsection{Pervasive Service-Based Architecture}

3GPP successfully introduced the Service-Based Architecture (SBA) to the 5G Core, replacing point-to-point, hard-coded network interfaces with a cloud-native, microservices-based model where Network Functions (NFs) communicate via RESTful APIs \cite{SBA-5G-2023}. The 6G architecture comprehensively expands the SBA paradigm and pushes the service-based design out of the core and aggressively into the Radio Access Network (RAN), disaggregating tightly coupled RAN protocols into modular, service-based components \cite{6G-Arch-23}, \cite{SBA-6G-RAN-25}. 

This pervasive SBA fulfills the service-orientation prerequisite of the SOVA framework. By abstracting network operations into highly granular, accessible microservices, the 6G infrastructure allows the middle layer of the SOVA framework to invoke precise network capabilities. Orchestration mechanisms can utilize abstract service publication and discovery to compose hybrid protocol bundles. 
The disaggregation promoted by alliances such as O-RAN further enables service-based components to run on generalized, vendor-neutral hardware, reducing coupling and enhancing flexibility of the physical network \cite{6G-AI-edge-25}. 

\subsection{AI-as-a-Serivce}

A key aspect of the transition to AI-nativeness in the 6G network is reflected in the AI-as-a-Service (AIaaS) paradigm, listed as a key issue (KI\# 19) ``6G Network for AI'' in 3GPP TR 23.801 \cite{TR-23-801}. 
3GPP is actively studying how the 6G Core can serve as a secure, high-performance broker for AI services toward realizing the AIaaS paradigm. TR 23.801 mandates that the 6G network natively support AI Agent Discovery, enabling an autonomous AI agent on a mobile device to effortlessly locate and query specialized AI agents hosted on distant edge servers. AIaaS demands that the network facilitate highly secure, authenticated agent-to-agent communications and explicitly expose deep network capabilities to these AI agents via robust, standardized APIs. 

The emerging AIaaS paradigm in the 6G network aligns flawlessly with the SOVA framework's Agent-as-a-Service concept; therefore, AIaaS in 6G offers a promising approach to achieving the SOVA framework’s objective of abstracting agent capabilities into manageable, scalable, and composable platform services. The AIaaS paradigm, together with the SBA in the 6G network, may provide effective mechanisms for implementing cross-domain discovery and orchestration of agent services and network services, which are critical functionalities required by the SOVA framework for AI agent communications.

\subsection{Intent-Based Networking}

The AI-native design principle of the 6G network is also evident in the evolution of the network analytics architecture. In 5G, the Network Data Analytics Function (NWDAF) was introduced primarily as a passive observer that collects data and provides analytics to other network functions \cite{duan2021intelligent}. The 6G architecture introduces an enhanced NWDAF (eNWDAF) comprising dedicated AI agents embedded in both the core and the radio edge \cite{eNWDAF-24}, which are responsible for managing the lifecycle of AI operations within the network, including localized model training, real-time inference, and the execution of closed-loop control policies across the network topology \cite{LLM-NWDAF-26}. 

This embedded intelligence directly enables Intent-Based Networking (IBN), allowing 6G to leverage LLMs and reinforcement learning methods to abstract the complexity of network management \cite{IBN-6G-26}. Instead of highly trained human network operators manually configuring routing protocols or QoS priorities, end users or overlying agentic AI applications simply provide high-level, declarative intents in natural language. The IBN capabilities in the 6G network may greatly facilitate the inter-layer cooperation in the SOVA framework by translating abstract semantic intents into concrete, multi-domain network configurations via LLM reasoning without human intervention. 

\subsection{Semantic Communications}

Semantic communication (SemCom) is expected to play a crucial role in the 6G network. Early SemCom relied on deep joint source-channel coding (JSCC) to reconstruct imagery under channel impairments; modern semantic-agentic frameworks utilize knowledge graphs and large language models (LLMs) to infer goals and translate them into optimal transport protocols \cite{SemCom-survey-25}. SemCom in the 6G network enables the physical layer to drastically compress payloads based on their semantics, conserving critical edge battery power, preserving limited radio spectrum, and dramatically accelerating communication speeds \cite{SemCom-6G-2025}. 

For autonomous agentic communications, where LLM-based AI agents frequently generate highly structured but massively redundant metadata (such as extremely verbose JSON-LD semantic descriptions or repetitive API schemas). SemCom necessitates moving away from traditional Quality of Service (QoS) metrics -- which obsess over raw throughput, latency, and bit-error rates -- toward a new framework of Semantic QoS that rigorously measures the accuracy of task completion, the preservation of contextual intent, and semantic fidelity at the receiver end \cite{SemCom-6G-26}, which plays a key role in high-performance communications for multi-agent collaborations.  

\begin{table}[htbp]
\centering
\caption{Alignment of AI-native 6G architectural capabilities and the SOVA framework for AI agent communications}
\label{tab:6g_capabilities}
\footnotesize
\renewcommand{\arraystretch}{1.2}
\begin{tabular}{|p{1.6cm}|p{3.0cm}|p{3.0cm}|}
\hline
\textbf{6G Attributes} & \textbf{6G Capabilities} & \textbf{Support of SOVA for Agent Communications} \\ \hline
Computing-Networking Slicing & dynamic, simultaneous allocation of network and compute resources & realizes the virtualized infrastructure; provides precise resources tailored to agent protocols \\ \hline
Pervasive Service-Based Architecture & extension of the service-oriented architecture from Core into RAN & fulfills service orientation; allows orchestration of network behaviors by MAS platforms \\ \hline
AI-as-a-Service & embedded control-plane network functions for AI agent discovery, hosting, and exposure & matches the Agent-as-a-Service concept in SOVA; enables native, efficient cross-domain agent discovery \\ \hline
Intent-Based Networking & autonomous, closed-loop network configuration driven by declarative natural language or JSON intents & bridges application-layer MAS requirements with network infrastructure operations \\ \hline
Semantic Communications & prioritizing transmission of meaning over raw bit accuracy & achieving efficient information exchange between AI agents by transmitting only semantic meaning \\ \hline
\end{tabular}
\end{table}

\section{Gap Analysis: Current 6G Specifications vs. AI Agent Communications}

While the developmental trajectory of 6G technology closely aligns with the demanding requirements of the Internet of AI Agents, a critical analysis of current specifications (including the early Release 19 and Release 20 drafts) reveals several significant gaps. These gaps exist precisely at the intersection where the demands of the SOVA framework for agent communication meet the current and developing capabilities of the 6G network infrastructure. 

\subsection{The Infrastructure Gap: Deficient Cross-Layer Service Orchestration}

The SOVA framework inherently requires an integrated cross-layer architecture in which the Communication Platform Layer (actively executing agent protocols such as the merged A2A+ACP protocol) interacts seamlessly and bidirectionally with the Virtualized Infrastructure Layer (the underlying 6G network slices).
However, currently, 6G SBA and O-RAN orchestration frameworks operate in deeply siloed domains, focusing on intra-network optimization (e.g., using AI to optimize radio spectrum or predict handovers) rather than true cross-layer, application-to-network co-optimization. While TR 23.801 (KI\#19) boldly introduces the concept of AIaaS and network capability exposure, the specific signaling mechanisms required for a multi-agent system to communicate its real-time, fluctuating resource constraints down to the 6G control plane remain undefined. Consider a dynamic scenario in which a collaborative swarm of AI agents suddenly shifts from asynchronous, background data gathering (which easily tolerates high delays) to a real-time, synchronous consensus mechanism that requires instantaneous communication. The SOVA platform must immediately request a dynamic slice reconfiguration to a URLLC profile. However, existing 3GPP Network Exposure Functions (NEFs) lack the low-latency, fine-grained, semantic-aware APIs required for an external AI agent to successfully negotiate sub-second Compute-Network resource reallocation. 

\subsection{The Interoperability Gap: Semantic Alignment and Processing Overhead}

Achieving true interoperability across the Internet of AI Agents requires highly robust, deeply expressive agent-description mechanisms. Advanced protocols like ANP and LMOS rely heavily on JSON-LD and Web of Things (WoT) Thing Descriptions to impart meaning to data, while the AGNTCY protocol utilizes the Open Agent Schema Framework (OASF) to embed vital computational context (e.g., current CPU load, battery life) alongside an agent's functional skills \cite{OCI-spec}. 
However, the 6G specifications have yet to standardize a universal ontology, vocabulary, or semantic schema for defining AI agents as recognizable network entities. While TR 23.801 actively discusses Agent Discovery, if the underlying 6G Core Network cannot natively parse these complex semantic capability descriptions, the network cannot effectively perform its brokering duties. For instance, the network must be able to instantly distinguish between a lightweight text-processing agent that requires minimal resources and a heavy vision-processing agent that requires a massive allocation of edge GPU resources. Furthermore, resolving massive JSON-LD linked data graphs incurs staggering computational penalties. Current 6G networks lack a lightweight, standardized metadata protocol layer designed to cache, rapidly parse, and route these highly heterogeneous semantic descriptions without triggering catastrophic signaling storms within the core control. 

\subsection{The Slicing Gap: QoS-Driven Slicing vs. True Semantic Slicing}

A key requirement of the SOVA framework for agent communications is the ability to compose and provision network services tailored to highly diverse multi-agent interactions. 
However, 6G network slicing, as currently specified, remains largely QoS-driven, focusing on rigid metrics such as maximum bit rate, latency bounds, and jitter tolerances. Efficient agent communication requires implementing Semantic Slicing, which dynamically allocates physical network resources based on the data flow's actual meaning and specific contextual task, rather than just its raw volumetric requirements \cite{Semantic-Slicing-25}. For example, if an AI agent uses a high-definition video feed exclusively to detect specific anomalies on a factory floor, the agent does not require a pristine 4K video stream (which would consume a massive eMBB slice). It requires only that the specific semantic features of the anomaly be transmitted intact. Current 3GPP management and orchestration (MANO) standards, such as TS 28.530 \cite{TS-28-530}, do not inherently incorporate semantic QoS thresholds into their slice selection algorithms. The absence of an AI-native semantic-based dynamic slicing framework forces the network into severe over-provisioning, preventing the SOVA framework from efficiently executing agent communications in resource-constrained edge environments.

\subsection{The Security Gap: Centralized Authority vs. Decentralized Trust}

The current 3GPP 6G network security architecture, including the latest iterations in TR 33.801 \cite{TR-33-801}, is anchored in Central Authority (CA) models that rely on hardware SIM credentials, the centralized Authentication Server Function (AUSF), and Unified Data Management (UDM) databases. While highly effective for traditional, human-operated mobile networks, this centralized model becomes a bottleneck and a single point of failure for the decentralized Internet of AI Agents. Emerging agent communication protocols (such as ANP and AGNTCY) strongly favor Decentralized Identifiers (DIDs) and cryptographically secure Verifiable Credentials (VCs) \cite{DID-survey-2025} to enable trustless, peer-to-peer authentication across diverse enterprise domains without relying on a central server. Current 6G specifications do not natively bridge 3GPP cellular hardware identities with W3C self-sovereign DIDs. Consequently, the SOVA framework for agent communication is forced to run heavy application-layer overlay security networks, which drastically increase latency and overhead.

\subsection{The Efficiency Gap: Agentic Flexibility vs. Resource-Constrained Mobile Networks}

To successfully address the heterogeneity challenge of agent communication, the SOVA framework relies on intelligent, adaptive protocol negotiation. It must seamlessly shift between RESTful JSON, standard JSON-RPC, and compressed binary gRPC (SLIM) based on the immediate capabilities of the deployment environment.
However, there exists an efficiency mismatch between this SOVA flexibility and the 6G network infrastructure. While 6G networks support all these application-layer protocols transparently as simple data payloads, running enterprise-grade protocols in mobile edge environments is devastatingly inefficient. Protocols like A2A, which heavily use JSON-RPC over HTTP/HTTPS, incur significant computational and serialization overhead. Text-based JSON serialization results in 30\% to 50\% larger payloads than binary formats, consuming excessive computing resources on 6G mobile edge devices. Conversely, while AGNTCY's SLIM messaging scheme utilizes gRPC over HTTP/2 to achieve high throughput and low latency, it inherently assumes highly stable, uninterrupted TCP connections, which may not be feasible in high-mobility 6G scenarios. The 6G Core currently lacks native, intelligent middleware that automatically transcodes or proxies these heavy application protocols into ultra-lightweight, edge-native formats (such as CoAP or MQTT-SN) in response to fluctuating radio channel state information (CSI).

\begin{table}[t]
\centering
\caption{Gap analysis between current 6G specification and AI Agent communication protocol requirements}
\label{tab:gap_analysis}
\footnotesize
\renewcommand{\arraystretch}{1.2}
\begin{tabular}{|p{1.4cm}|p{2.8cm}|p{3.0cm}|}
\hline
\textbf{Gap Category} & \textbf{Current 6G Specification Status} & \textbf{AI Agent Protocol Requirement} \\ \hline
Infrastructure Orchestration & siloed intra-network AI optimization; generic NEF API exposure for external entities & real-time, continuous cross-layer resource negotiation driven by the SOVA framework \\ \hline
Semantic Interoperability & lack of a universal metadata ontology for AI agents within the 6G core network & JSON-LD, WoT Thing Descriptions (LMOS), and OASF complex capabilities schemas \\ \hline
Dynamic Slicing & QoS-driven parameter matching based on raw volume, jitter, and latency limits & Semantic Slicing based on intended meaning, context, and required task resolution \\ \hline
Security \& Trust & hardware-rooted centralized authority utilizing legacy AKA frameworks & decentralized identifiers and trustless peer-to-peer verification \\ \hline
Protocol Efficiency & transparent IP transport of any protocol, agnostic to edge compute suitability & adaptive meta-protocol negotiation; capable of handling high-mobility packet loss \\ \hline
\end{tabular}
\end{table}

\subsection{The Synchronization Gap: 6G Specifications vs. Agent Protocols}

The 6G standardization has explicitly identified AI agents as foundational components of the network. 
3GPP TR 22.870 \cite{TR-22-870} mandates stringent requirements for agent interoperability (including standardized protocols and multimodal data formats), dynamic discovery mechanisms across domains, robust task management (decomposition, scheduling, and monitoring), and deep context awareness. 
However, 3GPP fundamentally defines service requirements and system architectures; it does not specify the application-layer communication protocols themselves. The development of intent-based networking and semantic agent protocols largely falls under the purview of organizations such as the Internet Engineering Task Force (IETF) and the World Wide Web Consortium (W3C), as well as open-source consortia (e.g., the Linux Foundation, the Eclipse Foundation) that drive agent communication protocols such as A2A, ACP, and LMOS. Currently, a synchronization gap exists between these bodies. The emerging software-driven agent protocols are being developed with inadequate consideration for the deterministic QoS, mobility management, and radio resource control paradigms defined by 3GPP. Without active, formalized coordination between 3GPP and the IETF/W3C/open-source consortia, the resulting AI agent communication protocols will fail to natively interface with the 6G network control plane, leaving the infrastructure gap unresolved.

\section{Possible Research Directions for AI Agent Communications in AI-Native 6G Network} 

This section outlines potential research and development directions to ensure that the AI-native 6G network specifications seamlessly support AI agent communications toward realizing the Internet of AI agents. 

\subsection{Standardizing Semantic QoS and Enforcing Semantic Network Slicing}

To overcome the severe inefficiencies associated with transmitting verbose semantic agent descriptions (like JSON-LD) and heavy agent-to-agent data payloads, the 6G architecture needs to fully embrace Semantic Communications. 
Future research should prioritize developing a novel core network entity: the Semantic Slice Management Function (SSMF). This function would reside within the 6G Core and be capable of deeply inspecting and interpreting the semantic context and ultimate intent of an AI agent's data flow. The SSMF would dynamically allocate compute and radio resources based strictly on a Semantic QoS threshold. This threshold represents the absolute minimum feature extraction required for the receiving agent to successfully complete its assigned task, thereby preventing massive network over-provisioning. Furthermore, intensive investigation is required to integrate Semantic Communication mechanisms directly into standard agent communication protocols (such as the merged A2A or ACP). By training advanced autoencoders to compress agent intents and verbose JSON schemas based on real-time, fluctuating 6G channel conditions, the network can transmit highly compressed semantic embeddings rather than raw ASCII text. This will dramatically reduce power consumption, conserve spectrum, and minimize latency at the extreme edge.

\subsection{Agentic Intent-Based Networking for Cross-Layer Orchestration}

Bridging the critical infrastructure gap requires the SOVA Communication Platform Layer to govern the 6G Virtualized Infrastructure Layer seamlessly and bidirectionally, without violating strict network security constraints. 
Research efforts should focus on deploying highly specialized LLM agents (designated as Core Agents and RAN Agents) directly within the 6G management plane. These embedded agents will be designed to ingest natural language commands or highly formalized XML/JSON intents from the external SOVA platform. Utilizing iterative Reason+Act (ReAct) prompting cycles, these agents will automatically synthesize, test, and execute the highly complex API calls required to dynamically scale virtual machines, adjust MIMO beamforming parameters, or allocate new URLLC slices \cite{IBN-6G-26}. To facilitate this, the industry needs to standardize a universal Intent-Translation API schema. This schema will allow external, third-party MAS platforms to express broad operational intents (e.g., ``maximize data gathering speed for this distributed sensor swarm while minimizing battery drain'') to the 6G network without requiring the AI application developers to possess any underlying knowledge of complex 3GPP signaling protocols.

\subsection{Evolving eNWDAF into a Distributed Agentic Service Broker}

While 3GPP TR 23.801 (KI\#19) boldly opens the door for AI-as-a-Service in the 6G network, the network must evolve beyond simple DNS-like service discovery and move toward active, intelligent service mediation.  
Research is needed to drive the transition of the enhanced NWDAF (eNWDAF) into a fully distributed, intelligent agentic broker. When a mobile client agent on a UE seeks a specific, computationally heavy service, the eNWDAF should not merely provide a list of IP addresses of available cloud agents. Instead, utilizing its omniscient global view of network congestion, radio interference, and real-time Compute-Network Convergence (CNC) states, the eNWDAF must actively recommend and route the request to the optimal agent that minimizes end-to-end latency and compute cost. To support resource-aware agent discovery without flooding the core control plane with signaling storms, researchers can apply the OASF-inspired ``incremental metadata update'' principle directly within the 6G Network Repository Function (NRF) or the eNWDAF itself. In this paradigm, edge agents would transmit only differential updates for highly dynamic states—such as current battery percentage, GPU availability, or physical location—rather than repeatedly transmitting their full semantic manifests.

\subsection{Edge-Native Protocol Transcoding via Service-Based RAN} 

The severe inefficiency of running enterprise-grade protocols (such as JSON-RPC over HTTP) in resource-constrained, dynamic edge environments must be mitigated directly at the radio access edge to support large-scale deployment of AI agents on edge devices. 
Future network architectures should be developed with a strict protocol-agnostic philosophy, integrating active protocol transcoders as dynamic microservices directly within the Service-based RAN. For example, if a massive cloud-native research agent communicates via high-speed gRPC (SLIM), but the receiving edge sensor agent operates on a low-power mMTC slice using a lightweight protocol like CoAP, the Service-based RAN should autonomously detect this mismatch. It should then instantly instantiate a virtualized proxy at the cell tower to seamlessly translate the payload format and shift the message-exchange pattern (e.g., from synchronous streaming to asynchronous queueing) in real time. Additionally, leveraging LLMs for semantic parsing of RAN services will enable dynamic, on-the-fly recomposition of atomic network functions (e.g., MAC scheduling algorithms, PHY encoding schemes) specifically tuned to perfectly accommodate the highly bursty, hyper-latency-sensitive nature of multi-agent inference traffic \cite{AI-PHY-6G-25}. 

\subsection{Hybrid Trust and Zero-Trust Architectures (ZTA)} 

To fully secure the vast, decentralized Internet of AI agents against novel threats, network security paradigms must transcend traditional physical network perimeters and operate actively at the semantic layer.  
A paramount research direction is the development of Hybrid Authentication Protocols. These protocols must build seamless, ultra-fast cryptographic bridges between 3GPP's centralized Authentication and Key Agreement (AKA) frameworks and the decentralized W3C DID/VC standards used by advanced AI protocols such as ANP and AGNTCY. A 6G edge node (e.g., a next-generation gNB or UPF) could be engineered to act as a highly trusted, localized DID resolver. Utilizing specialized cryptographic hardware acceleration, this edge node could verify decentralized credentials locally within microseconds. This architecture would maintain strict URLLC latency budgets while fully enabling decentralized, trustless agent interactions across disparate enterprise domains. Furthermore, because traditional firewalls are completely blind to semantic attacks such as prompt injection hidden within legitimate JSON payloads, research and development should focus on deploying Semantic Intrusion Detection Systems (S-IDS). This involves deploying lightweight, highly quantized LLMs directly at the 6G edge (e.g., embedded within the UPF or eNWDAF) to perform continuous, real-time semantic behavioral monitoring. These AI monitors would establish deep behavioral baselines for every registered agent and possess the authority to instantly revoke network slice resources if an agent suddenly exhibits anomalous behavior indicative of hijacked intents.

\section*{Conclusions}
In this paper, we comprehensively assessed the status, challenges, and opportunities of supporting AI agent communications within the emerging AI-native 6G network. We first discussed the diverse landscape of current agent communication protocols and their inherent limitations, highlighting the proposed Service-Oriented Virtualization-Based Architecture (SOVA) framework as a unified approach to resolving interoperability and infrastructure gaps. We then focused our investigation on evaluating core 6G design principles, demonstrating how capabilities such as advanced network slicing, pervasive Service-Based Architecture, the AI-as-a-Service paradigm, Intent-Based Networking, and Semantic Communications natively align with the requirements of the SOVA framework. Our critical analysis of current 6G specifications indicates that architectural and protocol-level gaps remain, particularly regarding cross-layer service orchestration, overheads for processing semantic schemas and agents' intentions, and the lack of semantic slicing and decentralized trust frameworks. Based on these insights, we outlined key directions for future research and development to bridge the gap in synchronization between agent protocol development and 6G standardization toward realizing the Internet of AI agents upon the global telecommunications infrastructure.

\bibliographystyle{ieeetr}
\bibliography{AgentComm-6G-Ref}

\begin{thebibliography}{10}

\bibitem{Rise-Agent-survey-2025}
Z.~Xi, W.~Chen, X.~Guo, W.~He, Y.~Ding, B.~Hong, M.~Zhang, J.~Wang, S.~Jin, E.~Zhou, {\em et~al.}, ``The rise and potential of large language model based agents: A survey,'' {\em Science China Information Sciences}, vol.~68, no.~2, p.~121101, 2025.

\bibitem{Agentic-AI-2026}
R.~Sapkota, K.~I. Roumeliotis, and M.~Karkee, ``{AI agents vs. agentic AI: A conceptual taxonomy, applications and challenges},'' {\em Information Fusion}, vol.~126, p.~103599, February 2026.

\bibitem{Self-talk-2025}
B.~Yan, Z.~Zhou, L.~Zhang, L.~Zhang, Z.~Zhou, D.~Miao, Z.~Li, C.~Li, and X.~Zhang, ``Beyond self-talk: A communication-centric survey of {LLM}-based multi-agent systems,'' {\em arXiv preprint arXiv:2502.14321}, 2025.

\bibitem{duan2025agent}
Q.~Duan and Z.~Lu, ``Agent communications toward agentic {AI} at edge-a case study of the {Agent2Agent} protocol,'' in {\em 2025 IEEE 11th International Conference on Edge Computing and Scalable Cloud (EdgeCom)}, pp.~150--155, November 2025.

\bibitem{Agent-Interoper-Survey-2025}
A.~Ehtesham, A.~Singh, G.~K. Gupta, and S.~Kumar, ``A survey of agent interoperability protocols: Model context protocol ({MCP}), agent communication protocol ({ACP}), agent-to-agent protocol ({A2A}), and agent network protocol ({ANP}),'' {\em arXiv preprint arXiv:2505.02279}, 2025.

\bibitem{Agent-Comm-Survey-2025}
D.~Kong, S.~Lin, Z.~Xu, Z.~Wang, M.~Li, Y.~Li, Y.~Zhang, Z.~Sha, Y.~Li, C.~Lin, {\em et~al.}, ``A survey of {LLM}-driven {AI} agent communication: Protocols, security risks, and defense countermeasures,'' {\em arXiv preprint arXiv:2506.19676}, 2025.

\bibitem{duan2026agent}
Q.~Duan and Z.~Lu, ``Ai agent communications in the future {Internet}—paving a path toward the agentic {Web},'' {\em Future Internet}, vol.~18, no.~3, p.~171, 2026.

\bibitem{AI-Comm-6G-25}
Q.~Cui, X.~You, N.~Wei, G.~Nan, X.~Zhang, J.~Zhang, X.~Lyu, M.~Ai, X.~Tao, Z.~Feng, {\em et~al.}, ``Overview of {AI} and communication for {6G} network: Fundamentals, challenges, and future research opportunities,'' {\em Science China Information Sciences}, vol.~68, no.~7, p.~171301, 2025.

\bibitem{A2A-Review-2025}
P.~P. Ray, ``A review on agent-to-agent protocol: concept, state-of-the-art, challenges and future directions,'' {\em Authorea Preprints}, 2025.

\bibitem{Secure-A2A-2025}
I.~Habler, K.~Huang, V.~S. Narajala, and P.~Kulkarni, ``Building a secure agentic {AI} application leveraging {A2A} protocol,'' {\em arXiv preprint arXiv:2504.16902}, 2025.

\bibitem{ACP-2026}
``{Agent Communication Protocol (ACP)}.'' \url{https://agentcommunicationprotocol.dev/introduction/welcome}.
\newblock Accessed: 2026-7-20.

\bibitem{ANP-WP-2025}
G.~Chang, E.~Lin, C.~Yuan, R.~Cai, B.~Chen, X.~Xie, and Y.~Zhang, ``Agent network protocol technical white paper,'' {\em arXiv preprint arXiv:2508.00007}, July 2025.

\bibitem{DID-survey-2025}
C.~Mazzocca, A.~Acar, S.~Uluagac, R.~Montanari, P.~Bellavista, and M.~Conti, ``A survey on decentralized identifiers and verifiable credentials,'' {\em IEEE Communications Surveys \& Tutorials}, vol.~27, pp.~3641 -- 3671, December 2025.

\bibitem{JSON-LD-spec}
W3C, ``{JSON-LD: A JSON-based Serialization for Linked Data version 1.1}.'' \url{https://www.w3.org/TR/json-ld11/}, July 2022.
\newblock Accessed: 2026-7-20.

\bibitem{LMOS}
Eclipse, ``{Language Model Operating System (LMOS)}.'' \url{https://eclipse.dev/lmos/}.
\newblock Accessed: 2026-7-20.

\bibitem{MQTT-spec}
OASIS, ``{Message Queuing Telemetry Transport (MQTT) version 5.0}.'' \url{https://www.oasis-open.org/standard/mqtt-v5-0-os/}, March 2019.
\newblock Accessed: 2026-7-20.

\bibitem{CoAP-spec}
IETF, ``{RFC 7252: The Constrained Application Protocol (CoAP)}.'' \url{https://datatracker.ietf.org/doc/html/rfc7252}, June 2014.
\newblock Accessed: 2026-7-20.

\bibitem{AGNTCY-origins}
AGNTCY/docs, ``{AGNTCY Origins}.'' \url{https://docs.agntcy.org/}, 2025.
\newblock Accessed: 2026-7-20.

\bibitem{Performance-gRPC-2024}
M.~Niswar, R.~A. Safruddin, A.~Bustamin, and I.~Aswad, ``Performance evaluation of microservices communication with rest, graphql, and grpc,'' {\em International Journal of Electronics and Telecommunication}, vol.~70, no.~2, pp.~429--436, 2024.

\bibitem{AI-native-6G-review-2025}
F.~C. Ogenyi, C.~N. Ugwu, and O.~P.-C. Ugwu, ``A comprehensive review of {AI-native 6G}: integrating semantic communications, reconfigurable intelligent surfaces, and edge intelligence for next-generation connectivity,'' {\em Frontiers in Communications and Networks}, vol.~6, p.~1655410, 2025.

\bibitem{NetSlice-5G-2019}
S.~Zhang, ``An overview of network slicing for {5G},'' {\em IEEE Wireless Communications}, vol.~26, no.~3, pp.~111--117, 2019.

\bibitem{NetSlice-6G-2022}
W.~Wu, C.~Zhou, M.~Li, H.~Wu, H.~Zhou, N.~Zhang, X.~S. Shen, and W.~Zhuang, ``{AI}-native network slicing for {6G} networks,'' {\em IEEE Wireless Communications}, vol.~29, no.~1, pp.~96--103, 2022.

\bibitem{duan2020convergence}
Q.~Duan, S.~Wang, and N.~Ansari, ``Convergence of networking and cloud/edge computing: Status, challenges, and opportunities,'' {\em IEEE Network}, vol.~34, no.~6, pp.~148--155, 2020.

\bibitem{CNC-6G-26}
Y.~Li, X.~Zhang, Y.~Zhang, and W.~Wang, ``Towards ubiquitous {6G} computing and networking convergence: Architecture and mechanism for cross-domain resource coordination,'' {\em arXiv preprint arXiv:2606.15073}, 2026.

\bibitem{Joint-NS-24}
Z.~Sasan, M.~Shokrnezhad, S.~Khorsandi, and T.~Taleb, ``Joint network slicing, routing, and in-network computing for energy-efficient {6G},'' in {\em 2024 IEEE Wireless Communications and Networking Conference (WCNC)}, pp.~1--6, 2024.

\bibitem{NS-VNF-26}
A.~C{\'a}rdenas, L.~Sigcha, and M.~Mosahebfard, ``Predictive network slicing resource orchestration: A {VNF} approach,'' {\em Future Internet}, vol.~18, no.~3, p.~149, 2026.

\bibitem{SBA-5G-2023}
K.~Du, L.~Wang, Z.~Zhu, Y.~Yan, and X.~Wen, ``Converged service-based architecture for next-generation mobile communication networks,'' in {\em 2023 IEEE Wireless Communications and Networking Conference (WCNC)}, pp.~1--6, 2023.

\bibitem{6G-Arch-23}
X.~D. Duan, X.~Y. Wang, L.~Lu, N.~X. Shi, C.~Liu, T.~Zhang, and T.~Sun, ``{6G} architecture design: From overall, logical and networking perspective,'' {\em IEEE communications magazine}, vol.~61, no.~7, pp.~158--164, 2023.

\bibitem{SBA-6G-RAN-25}
G.~Liu, N.~Li, C.~Yuan, S.~Chen, and X.~Liu, ``Service-based architecture for {6G RAN}: A cloud native platform that provides everything as a service,'' {\em Sensors}, vol.~25, no.~14, p.~4428, 2025.

\bibitem{6G-AI-edge-25}
C.~Feng, A.~Zhang, G.~Min, Y.~Huang, T.~Q. Quek, and X.~You, ``Towards {6G} native-{AI} edge networks: A semantic-aware and agentic intelligence paradigm,'' {\em arXiv preprint arXiv:2512.04405}, 2025.

\bibitem{TR-23-801}
3GPP, ``{TR 23.801 Study on Architecture for 6G System, Version 0.7.0},'' June 2026.
\newblock Accessed: 2026-7-10.

\bibitem{duan2021intelligent}
Q.~Duan, ``Intelligent and autonomous management in cloud-native future networks—a survey on related standards from an architectural perspective,'' {\em Future Internet}, vol.~13, no.~2, p.~42, 2021.

\bibitem{eNWDAF-24}
N.~V. de~Souza~Neto, M.~A. Gon{\c{c}}alves, D.~R.~C. Oliveira, D.~N. Molinos, R.~Moreira, and F.~de~Oliveira~Silva, ``Evolved {NWDAF} towards a fully distributed artificial intelligence in the {6G} network architecture,'' in {\em Workshop de Redes 6G (W6G)}, pp.~15--25, SBC, 2024.

\bibitem{LLM-NWDAF-26}
H.~Daniel, O.~Alhussein, C.~Li, J.~Liang, and E.~Damiani, ``{LLM}-enabled {NWDAF}: A step toward {AI}-native {6G} network intelligence,'' {\em https://doi.org/10.21203/rs.3.rs-8526650/v1}, 2026.

\bibitem{IBN-6G-26}
G.~Jiang, K.~Wang, X.~Chen, and Y.~Huang, ``Agentic {AI} empowered intent-based networking for 6g,'' {\em arXiv preprint arXiv:2601.06640}, 2026.

\bibitem{SemCom-survey-25}
T.~M. Getu, G.~Kaddoum, and M.~Bennis, ``Semantic communication: A survey on research landscape, challenges, and future directions,'' {\em Proceedings of the IEEE}, vol.~112, no.~11, pp.~1649--1685, 2025.

\bibitem{SemCom-6G-2025}
Y.~Wang, H.~Han, Y.~Feng, J.~Zheng, and B.~Zhang, ``Semantic communication empowered {6G} networks: Techniques, applications, and challenges,'' {\em IEEE Access}, vol.~13, pp.~28293--28314, 2025.

\bibitem{SemCom-6G-26}
S.~Sharif, F.~Khandaker, M.~Naeem, and W.~Ejaz, ``Resource optimization for semantic communication in {6G} networks: A survey,'' {\em IEEE Open Journal of the Communications Society}, vol.~7, February 2026.

\bibitem{OCI-spec}
OCI, ``{Open Container Initiative Runtime Specification}.'' \url{https://specs.opencontainers.org/runtime-spec/?v=v1.0.2}, November 2025.
\newblock Accessed: 2026-7-20.

\bibitem{Semantic-Slicing-25}
M.~R. Chowdhury, E.~Hammad, L.~Loven, S.~Pirttikangas, A.~P. Da~Silva, and W.~Saad, ``A framework for {AI}-native semantic-based dynamic slicing for {6G} networks,'' {\em arXiv preprint arXiv:2510.10756}, 2025.

\bibitem{TS-28-530}
3GPP, ``{TS 28.530 Management and orchestration; Concepts, use cases and requirements, Version 19.0.0},'' March 2025.
\newblock Accessed: 2026-7-10.

\bibitem{TR-33-801}
3GPP, ``{TR 33.801 Study on Security for the 6G System, Version 0.5.0},'' June 2026.
\newblock Accessed: 2026-7-10.

\bibitem{TR-22-870}
3GPP, ``{TR 22.870 Study on 6G Use Cases and Service Requirements, Version 20.0.0},'' March 2026.
\newblock Accessed: 2026-7-10.

\bibitem{AI-PHY-6G-25}
P.-M. Mutescu, A.-I. Petrariu, E.~Coca, C.~Patachia-Sultanoiu, R.~M. Mihai, and A.~Lavric, ``{AI}-native {PHY}-layer in {6G} orchestrated spectrum-aware networks,'' {\em Sensors}, vol.~25, no.~23, p.~7206, 2025.

\end{thebibliography}

\end{document}